\documentclass[a4paper,fleqn,usenatbib]{mnras}

\usepackage{newtxtext,newtxmath}

\usepackage[T1]{fontenc}
\usepackage{ae,aecompl}

\usepackage{graphicx}	
\usepackage{amsmath}	
\usepackage{amssymb}	

\title[Observations of CIZA J2242.8+5301 northern relic]{Spectro-polarimetric observations of the CIZA J2242.8+5301 northern radio relic: no evidence of high-frequency steepening}

\author[F. Loi et al.]{
F. Loi$^{1}$\thanks{E-mail: francesca.loi@inaf.it}, M. Murgia$^{1}$, V. Vacca$^{1}$, F. Govoni$^{1}$, A. Melis$^{1}$, D. Wittor$^{2}$,
R. Beck$^{3}$, 
\newauthor
M. Kierdorf$^{3}$, A. Bonafede$^{4,5}$, W. Boschin$^{6,7,8}$, M. Brienza$^{4,5}$,  E. Carretti$^{5}$, R. Concu$^{1}$, \newauthor
L. Feretti$^{5}$, F. Gastaldello$^{9}$, R. Paladino$^{5}$, K. Rajpurohit$^{4}$, P. Serra$^{1}$, and F. Vazza$^{2,4,5}$.\\
$^{1}$INAF-Osservatorio Astronomico di Cagliari, Via della Scienza 5, 09047 Selargius, IT\\
$^{2}$University of Hamburg, Hamburger Sternwarte, Gojenbergsweg 112, 21029 Hamburg, Germany\\
$^{3}$Max-Planck-Institut für Radioastronomie, Auf dem Hügel 69, 53121 Bonn, Germany \\
$^{4}$Dipartimento di Fisica e Astronomia, Università di Bologna, via P. Gobetti 93/2, 40129, Bologna, Italy\\
$^{5}$INAF-Istituto di Radioastronomia, via P. Gobetti 101, 40129 Bologna, Italy\\
$^{6}$Fundaci\'on G. Galilei - INAF (Telescopio Nazionale Galileo), Rambla J. A. Fern\'andez P\'erez 7, E-38712 Bre\~na Baja (La Palma), Spain\\
$^{7}$Instituto de Astrof\'{\i}sica de Canarias, C/V\'{\i}a L\'actea s/n, E-38205 La Laguna (Tenerife), Spain\\
$^{8}$Departamento de Astrof\'{\i}sica, Univ. de La Laguna, Av. del Astrof\'{\i}sico Francisco S\'anchez s/n, E-38205 La Laguna (Tenerife), Spain\\
$^{9}$INAF – IASF Milano, Via Corti 12, 20133 Milano, Italy}
\date{Accepted 2020 July 27. Received 2020 July 27; in original form 2020 June 04}

\pubyear{2020}

\begin{document}
\label{firstpage}
\pagerange{\pageref{firstpage}--\pageref{lastpage}}
\maketitle

\begin{abstract}
Observations of radio relics at very high frequency (>10\,GHz) can help to understand how particles age and are (re-)accelerated in galaxy cluster outskirts and how magnetic fields are amplified in these environments. In this work, we present new single-dish 18.6\,GHz Sardinia Radio Telescope and 14.25\,GHz Effelsberg observations of the well known northern radio relic of CIZA J2242.8+5301. We detected the relic which shows a length of $\sim$1.8\,Mpc and a flux density equal to $\rm S_{14.25\,GHz}=(9.5\pm3.9)\,mJy$ and $\rm S_{18.6\,GHz}=(7.67\pm0.90)\,mJy$ at 14.25\,GHz and 18.6\,GHz respectively. The resulting best-fit model of the relic spectrum from 145\,MHz to 18.6\,GHz is a power-law spectrum with spectral index $\alpha=1.12\pm0.03$: no evidence of steepening has been found in the new data presented in this work. For the first time, polarisation properties have been derived at 18.6 GHz, revealing an averaged polarisation fraction of $\sim40\%$ and a magnetic field aligned with the 'filaments' or 'sheets' of the relic.
\end{abstract}




\section{Introduction}
The (re-)acceleration mechanism of relativistic particles in the intracluster medium of galaxy clusters is still poorly known as well as the origin of large-scale magnetic fields in such environments.\\
One of the spectacular manifestations of these components is represented by diffuse radio sources, known as radio halos and relics, respectively hosted at the centre and in the outskirts of galaxy clusters. These sources are faint (${\rm S_{\nu}\simeq 0.1-1\,\muup Jy/arcsec^{2}}$ at 1.4\,GHz) synchrotron sources, typically with steep power-law radio spectra (S$_{\nu}\sim\nu^{-\alpha}$, with $\alpha\sim$1), which extend over Mpc-scales \citep[see][for a recent review]{van19}. Radio relics are usually associated with shock waves propagating in the ICM as a consequence of galaxy cluster merging phenomena. This coincidence \citep{enss98} supports the diffusive shock acceleration \citep[DSA][]{drury,bland} as a mechanism of re-acceleration of cosmic-ray particles up to the $\sim$GeV energies required to explain the observed emission. 
According to DSA, particles scatter from magnetic field inhomogeneities and they cross the shock wave back and forth gaining energy at each crossing. The DSA mechanism generates a power-law energy distribution and, if the cooling of the particles is balanced by the injection of relativistic particles, a power-law behaviour of the flux density as a function of frequency. This trend has been observed in a large number of radio relics suggesting that these sources are produced by the DSA mechanism.
Other observed characteristics of radio relics are a gradual spectral index steepening toward the cluster centre, indicating that particles are ageing in the post-shock region, and a high degree of polarisation across the relic which suggests that the magnetic field has been compressed in a thin layer.
Indeed, shock compression can amplify the magnetic field component perpendicular to the shock direction. According to the models, the different mechanisms of magnetic field amplification can result in differences in the observed emission properties such as the radio spectrum, the surface brightness, and the spectral index profiles \citep[see for example the work of][]{donnert16}. \\
Several observations have challenged the DSA mechanism. These observations have revealed that for some relics the derived X-ray Mach numbers are low, namely weak shocks with M$\leq$3 \citep{brunetti,van19}. For such weak shocks, the DSA mechanism is not efficient enough to accelerate particles up to GeV energies from the thermal pool \citep[see also][]{botteon}. Only recently, a new class of radio relics with low surface brightness, and emissivity compatible with the standard DSA scenario might have been discovered at low frequency using LOFAR \citep{locatelli}.
In some cases, the Mach number inferred from X-ray observations and the one obtained from radio spectra are not in agreement \citep[see for example][]{van16}. In addition, some authors have reported the presence of a break in the spectrum, based on interferomteric observations, which is inconsistent with DSA \citep{stroe16,trasatti15}. \\
All these findings motivates the search for complementary/alternative models with respect to the DSA to explain the observed emission. 
In particular, it has been proposed that nearby active galactic nuclei can inject cosmic-ray electrons \citep{enss01}, a scenario confirmed in some cases \citep{bonafede14,van17,b19}, or that the injection is caused by powerful galactic wind \citep{volk}.
Spherically-expanding shocks with fossil particle populations \citep{kang_ciza} re-accelerated through DSA can generated curved spectra, while
a non-uniform magnetic field in the relic region \citep{donnert16} could cause a steepening in the radio spectrum.
Multiple shocks along the line-of-sight \citep[][and reference therein]{hong15} can explain the inconsistency between X-ray and radio derived Mach numbers. However, the modelling proposed by \citet{hong15} and \citet{donnert16} in the framework of the DSA does not solve the low efficiency problem which makes the acceleration of particles from the thermal pool unrealistic.\\
The inconsistency between the X-ray and radio derived Mach numbers \citep[see for example][for radio and X-ray derived Mach number]{van10,aka}, and the steepening in the flux density spectrum \citep{stroe16} was also observed for the northern relic of the galaxy cluster CIZA J2242.8+5301. 
This cluster (redshift z=0.1921) was discovered in the X-rays by \citet{koc07}, while its diffuse radio sources, a central radio halo and a pair of opposite radio relics, were discovered by \citet{van10}. 
Its northern relic is one of the most famous and extensively studied radio relics \citep{van10,stroe13,stroe14,stroe16,loi17,hoang17,kier17,digennaro18} and it is often considered a textbook example of radio relics because of its $\sim$2 Mpc arc-shaped morphology and uniform brightness along its length. \\
Interferometric measurements of the northern relic at 15.85\,GHz and 30\,GHz \citep{stroe16} obtained with the Arcminute Microkelvin Imager \citep[AMI,][]{ami} and Combined Array for Research in Millimeter-wave Astronomy \citep[CARMA,][]{carma} telescopes, triggered the search for alternative physics, as they indicated a steepening of the spectrum which is not compatible with the standard DSA model. Nevertheless, it should be noted that, especially at high frequency, interferometric observations can suffer from the missing zero baseline problem. While single-dish telescopes can retain angular structures as large as the observed area, interferometers can detect a maximum angular structure corresponding to their minimum baseline, since their minimum baseline is not equal to zero as in the case of single-dish. This means that interferometric observations are not able to fully recover the flux density of very extended sources. In particular, AMI small array (SA) and large array (LA) observations at 15.85\,GHz have a corresponding large angular scale $\theta_{LAS}\sim 11$\,arcmin and $\theta_{LAS}\sim 2$\,arcmin respectively. CARMA at 30\,GHz was blind to scales above $\theta_{LAS}\sim$3.5\,arcmin. The northern relic covers an angular scale larger than 11 arcmin. Therefore, even if \citet{stroe16} tried to correct for this effect, obtaining reliable flux density measurements from these telescopes is an hard task.\\
\citet{kier17} excluded a possible steepening between 153\,MHz and 8.35\,GHz and model the radio relic with a power-law with a spectral index $\alpha\sim$0.9 while DSA predicts $\alpha>1$.
In a previous work \citep{loi17}, we established, using data between $\sim$300\,MHz and $\sim$8\,GHz that the CIZA J2242.8+5301 northern relic spectral behaviour was consistent with the DSA model in this frequency range assuming a continuous injection of relativistic particles. From the spectral modelling, we also derived a Mach number consistent within the errors with the X-ray estimate \citep{aka}. However, the uncertainty about the relic behaviour at higher frequency remained.\\
In this work, we show the results of a large observing program (code: 72-19, P.I. Francesca Loi) conducted at the Sardinia Radio Telescope facility \citep[SRT,][]{bolli,prandoni} aiming at observing this famous relic at 18.6 GHz with the 7-feed K band receiver of this single-dish telescope in full-Stokes mode. We also present 14.25\,GHz data acquired with the Effelsberg single-dish telescope (code: 73-19, P.I. Rainer Beck).\\
In Sect. 2, we describe the observational set up, the data reduction and the imaging procedure for both total intensity and polarised intensity images.
In Sect. 3, we show the resulting total intensity image and the measurement of the relic flux, and compare these value with literature data. In Sect. 4, we discuss the Sunyaev-Zel'dovich effect \citep[SZ,][]{sz1,sz2} which could affect our measurements and based on the expected/observed contamination we give a rough estimate of the magnetic field in the relic region.
In Sect. 5, we show the polarised intensity image and we discuss the polarimetric properties of the detected sources.
In Sect. 6, we discuss our findings and draw the conclusions.\\
Throughout this paper, we assume a $\Lambda$CDM cosmology with ${\rm H_0 = 71\,km\cdot s^{-1}\cdot Mpc^{-1}}$ , ${\rm \Omega_m=0.27}$, and ${\rm \Omega_{\Lambda}=0.73}$. At the redshift of CIZA J2242.8+5301 (z=0.1921), 1 arcmin  corresponds to 189.9\,kpc.

\section{Observations and data reduction}
\subsection{SRT observations}
We used the 7-feed K-Band receiver of the SRT to observe at a central frequency of 18.6\,GHz inside a bandwidth of 1200\,MHz. These data were acquired with the SArdinia Roach2-based Digital Architecture for Radio Astronomy back end \citep[Sardara,][]{melis} using 1500 MHz bandwidth and 1024 channels of 1.46 MHz width in full-Stokes mode. \\
The observations have been carried out between January and April 2020, for a total of 240 hours divided in 27 slots of about 6-12 hours each. A summary of the observing program is reported in Table \ref{tab:obs}.
\begin{table}
	\centering
	\caption{Details about the SRT observations.}
	\label{tab:obs}
	\begin{tabular}{llll} 
		\hline
		Date & Obs. time & Calibrators & Number of maps\\
		\hline
		12 Jan 2020 & 8  hrs & 3C286, 3C84 & 8 RA + 8 Dec\\
        24 Jan 2020 & 7  hrs & 3C286, 3C84 & 11 RA + 10 Dec\\
        01 Feb 2020 & 8 hrs & 3C286, 3C84 & 10 RA + 10 Dec \\
        04 Feb 2020 & 10 hrs & 3C286, 3C84 & 11 RA + 11 Dec\\
        06 Feb 2020 & 8 hrs & 3C147, 3C84, 3C138 & 8 RA + 8 Dec \\
        07 Feb 2020 & 8 hrs & 3C286, 3C84 & 11 RA + 11 Dec \\
        08 Feb 2020 & 8 hrs & 3C286, 3C84 & 8 RA + 8 Dec  \\
        09 Feb 2020 & 8 hrs & 3C286, 3C84 & 11 RA + 11 Dec \\
        11 Feb 2020 & 11 hrs & 3C147, 3C84, 3C138 & 6 RA + 6 Dec\\
        12 Feb 2020 & 8 hrs & 3C286, 3C84 & 10 RA + 10 Dec \\
        26 Feb 2020 & 11 hrs & 3C286, 3C84 & 13 RA + 13 Dec \\
        27 Feb 2020 & 13 hrs & 3C286, 3C84 & 17 RA + 17 Dec \\
        28 Feb 2020 & 13 hrs & 3C286, 3C84 & 17 RA + 17 Dec \\
        29 Feb 2020 & 7 hrs & 3C286, NGC7027 & 10 RA + 10 Dec \\
        10 Mar 2020 & 6 hrs & 3C286, 3C84 & 9 RA + 9 Dec \\
        14 Mar 2020 & 8 hrs & 3C286, 3C84 & 12 RA + 12 Dec \\
        15 Mar 2020 & 8 hrs & 3C286, 3C84 & 11 RA + 11 Dec \\
        19 Mar 2020 & 10 hrs & 3C147, 3C84, 3C138 & 14 RA + 14 Dec\\
        24 Mar 2020 & 9 hrs & 3C147, 3C84, 3C138 & 11 RA + 11 Dec\\
        25 Mar 2020 & 6 hrs & 3C48, 3C147 & 8 RA + 8 Dec\\
        26 Mar 2020 & 7 hrs & 3C48, 3C84, 3C138 & 9 RA + 9 Dec \\
        02 Apr 2020 & 8 hrs & 3C286, 3C84 & 11 RA + 11 Dec \\
        03 Apr 2020 & 6 hrs & 3C48, 3C84, 3C138 & 5 RA + 5 Dec\\
        04 Apr 2020 & 10 hrs & 3C286, 3C84 & 16 RA + 16 Dec \\
        10 Apr 2020 & 8 hrs & 3C286, 3C84 & 12 RA + 14 Dec\\
        14 Apr 2020 & 11 hrs & 3C147, 3C84, 3C138 & 11 RA + 11 Dec\\
        16 Apr 2020 & 10 hrs & 3C286, 3C84 & 14 RA + 14 Dec \\
        \hline
	\end{tabular}
\end{table}
During each slot, we observed the primary calibrator 3C286 (or 3C147 if the former was not available) to calibrate band pass and flux density scale. We performed sky dips to derive the trend of the system temperature with elevation during our observations. We then modelled the T$_{\rm sys}$ trend with the airmass model to obtain the zenithal opacity, $\tau$. The values of $\tau$ derived from the sky dip were generally in good agreement with those provided by the radiometer working at the SRT site \citep{buffa}.
We also used the calibrator 3C286 as a reference for the absolute polarisation position angle \citep[we assumed the values from][]{pb2013}. If 3C286 was not available, we used 3C138 instead. The sources 3C84 and NGC7027, that we considered to be completely unpolarised, were used to correct for the on-axis instrumental polarisation. The northern relic of CIZA J2242.8+5301 was observed with the on-the-fly strategy in the equatorial frame, covering an area of 21$\times$15\,arcmin$^2$ centred on the relic centre (R.A. 22h 42m 58s, Dec. +53d 07m 12s). To facilitate the removal of the scan noise, we acquired orthogonal on-the-fly maps along the RA and DEC directions. Individual sub-scans within these maps are separated by 15 arcseconds, in order to sample the FWHM of the SRT beam with about 4 pixels along each direction. \\
We performed the data reduction and the imaging with the proprietary software package Single-dish Spectral-polarimetry Software \citep[SCUBE,][]{murgia}. We used a standardised pipeline for the calibration and imaging: we excised all the RFIs at well known specific frequencies and we applied an automatic flagging procedure to remove the remaining RFIs. We then determined the opacity, we subtracted the calibrator baselines with a linear fit of each scan based on the first and last 10\% of the data, we determined the bandpass solution, we gridded the calibrators in a regular grid with a pixel size of 15 arcsec, and we used the resulting images to fit a 2D-Gaussian to determine the flux scale and the leakage terms.\\
Our primary goal is to image the emission of the relic and the point sources in the field-of-view of target. Since we are uninterested to retain any large scale foreground emission, in the data of the target we removed the baseline scan-by-scan by fitting a 2nd order polynomial to the "cold-sky" regions devoid of both discrete sources and of the galaxy cluster extended emission (relics and halo). These cold-sky regions are identified using a mask created with the 1.4\,GHz SRT+WSRT image presented in \citet{loi17}, convolved with a beam FWHM of 1 arcmin. In this way, we removed the baselevel related to the receiver noise, the atmospheric emission, and the large scale foreground sky emission. 
We then imaged the spectral cubes using a regular grid of 15 arcsec of pixel size and we averaged all the spectral channel to increase the signal-to-noise ratio. The images from all the observing slots are stacked together to reduce the noise level. We stack the RA and DEC scans by mixing their stationary wavelet transform (SWT) coefficients \citep[see][]{murgia}. The de-stripping resulting from the SWT stacking is effective in isolating and removing the noise along the scan direction.
After the first stack was completed, we returned to the individual images using the higher signal-to-noise image as a reference model to flag all residual low-level RFIs or small-scale atmospheric fluctuations that were not captured at the flag and baseline removal stages. This refined flagging step significantly improved the quality of the images. In order to verify the consistency of the flux density scale calibration between all the different observing slots, we performed a self-calibration procedure using the cluster central source (R.A. 22h 42m 51.30s, Dec. +53d 04m 41.40s) as reference. We assumed that this source has a stable flux density of 8.8\,mJy and, by mean of a 2D Gaussian fit to this source, we calculated a correction factor for each observing slot. By analysing the results of the self-calibration procedure, we deduced that the rms scatter of the flux density calibration scale through our project is of about $\sim$10\%. This scatter could be assumed as an estimate for our systematic uncertainty of the flux density calibration.\\

The polarised image has been obtained after correcting for the leakage term determined with 3C84 (or NGC7027). We calibrated the polarised angle and fraction with 3C286 (or 3C138). To perform the imaging of the Q and U Stokes parameters, we followed the same steps described above for the total intensity. The images of the polarised intensity P and the polarisation angle $\Psi$ have been derived from the Stokes parameters solving for the following equations:
\begin{eqnarray}
\rm    P & = & \sqrt{\rm Q^2+U^2},\\ 
    \Psi & = & 0.5 \cdot \arctan{\rm \frac{U}{Q}} .
\end{eqnarray}
We also corrected the polarisation images for the Rician bias \citep[see][]{murgia}.\\
The maximum Rotation Measure observed for this relic \citep[see][]{loi17,kier17} is RM=-400\,rad\,m$^{-2}$. At 18.6\,GHz this would cause a rotation of the polarisation plane of $\Delta\Psi\sim$6\,deg. Therefore, it is unlikely that our signal can undergo significant depolarisation effect but it is indeed very close to the intrinsic polarisation properties of the relic.
\begin{figure*}
    \centering
    \includegraphics[width=1.9\columnwidth]{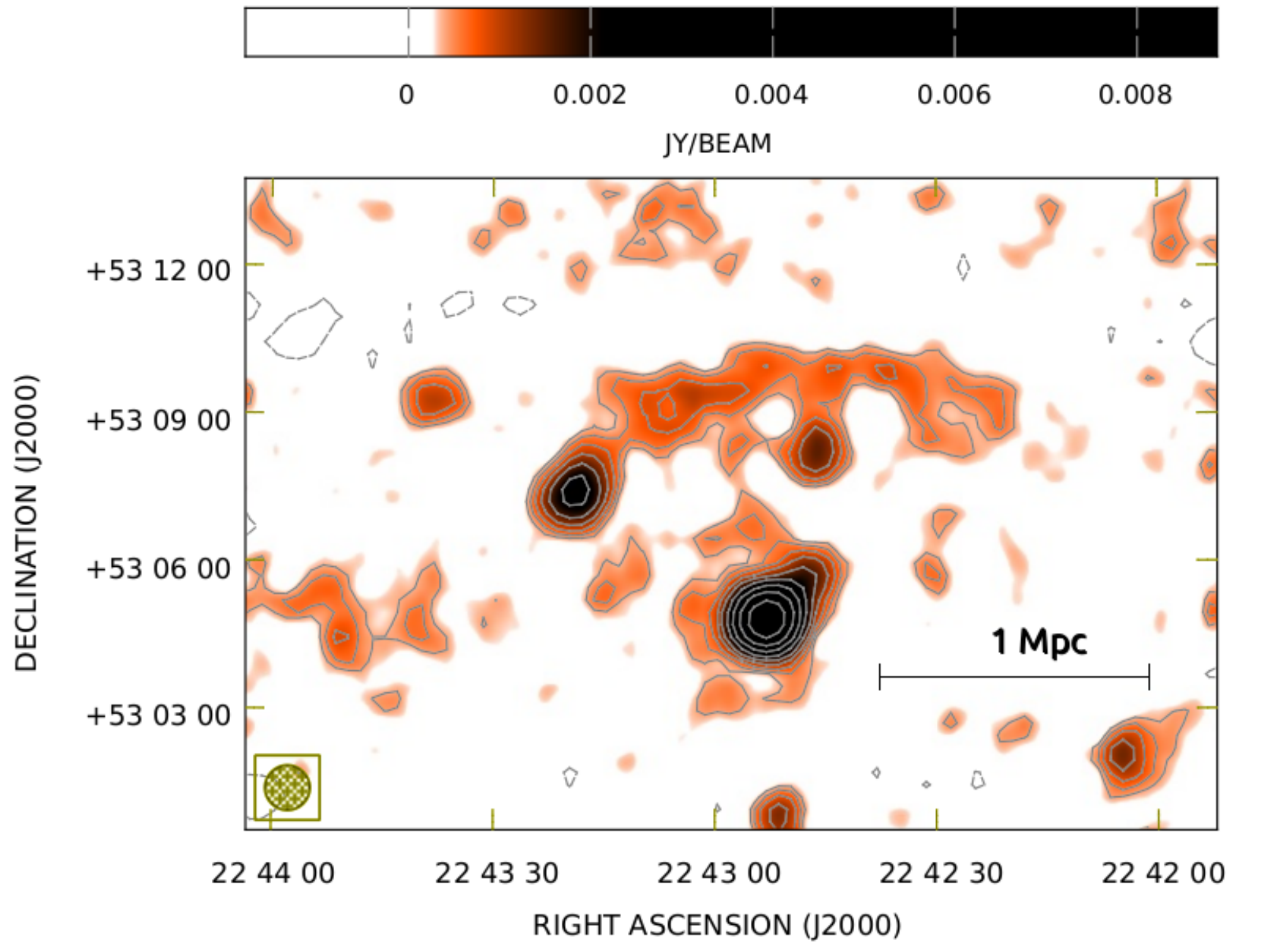}
    \caption{18.6\,GHz SRT total intensity image between 18\, and 19.2\,GHz. Contours start at 3$\sigma$ where $\sigma$=0.13\,mJy\,beam$^{-1}$ and increment with a $\sqrt{2}$ factor. Dashed contours are negative contours drawn at -3$\sigma$. The beam size is shown in the bottom left corner of the image, its FWHM corresponds to $\sim$0.9 arcmin.}
    \label{fig:ciza}
\end{figure*}
\begin{figure}
    \centering
    \includegraphics[width=\columnwidth]{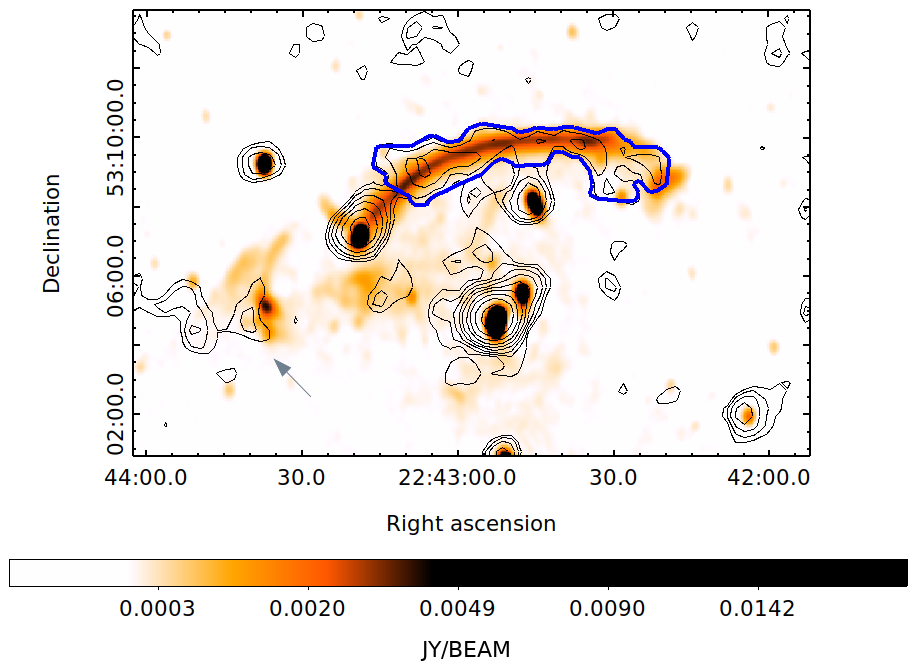}
    \caption{1.4\,GHz WSRT+SRT total intensity image from \citet{loi17} with the positive contours of Fig. \ref{fig:ciza}. The blue contour indicates the region where we estimated the flux density of the relic.}
    \label{fig:ciza_cnt_box}
\end{figure}
\subsection{Effelsberg observations}
We also observed the northern relic of CIZA J2242.8+5301 with the new two-horn Ku-band receiver of the Effelsberg 100-m telescope that provides two channels of 2.5\,GHz bandwidth each, centred on 14.25\,GHz and 16.75\,GHz. The system temperatures are 33\,K and 44\,K. In December 2019 and January 2020 we obtained 12 coverages of a field of 12$\times$9\,arcmin$^2$ by scanning ("on-the fly") in alternating RA and DEC directions. The total on-source observation time was 7 hours. Data processing (RFI removal, baselevel corrections, and combination of the coverages in RA and DEC with the "basket weaving" technique) was performed with the NOD3 software package \citep{nod} using 3C147 or 3C48 as flux density calibrators. The final image at 14.25 GHz with a resolution of 49\,arcsec has a rms noise of 1\,mJy\,beam$^{-1}$. To increase the signal-to-noise ratio, we smoothed the image to 72\,arcsec.

In the second channel centred at 16.75\,GHz no significant signal could be
detected. The second horn is separated in azimuthal direction by 3.85\,arcmin
and can be used to reduce weather effects. As this requires scanning larger
areas in azimuthal direction and the weather was excellent, this method was
not used. The linear polarisation signal was also recorded, but was too weak
due to the relatively short observation time.
\begin{figure*}
    \centering
    \includegraphics[width=1.5\columnwidth]{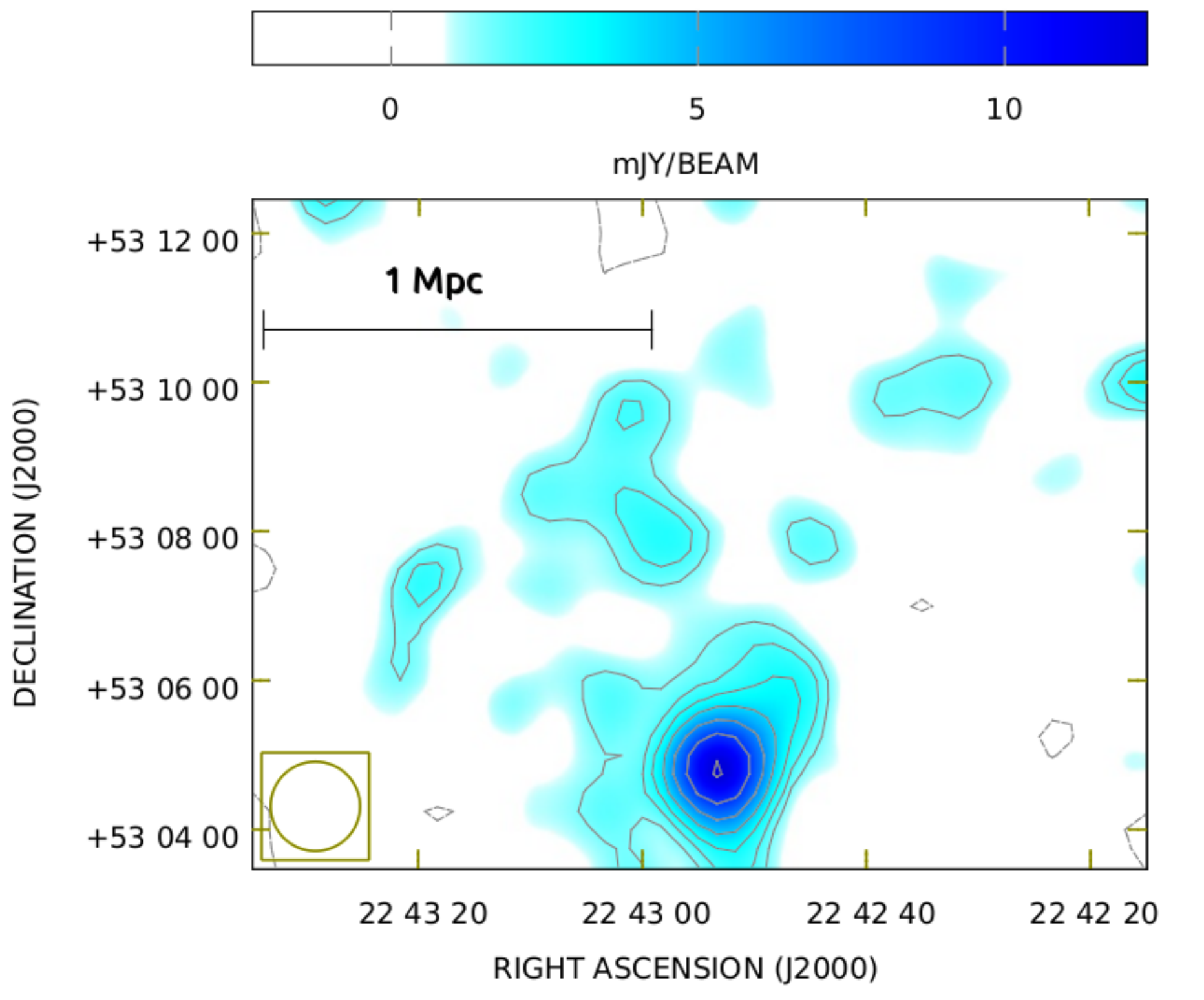}
    \caption{14.25\,GHz Effelsberg total intensity image between 13.0 and 15.5\,GHz. Positive contours (solid line) start at 3$\sigma$ where $\sigma=0.5$\,mJy\,beam$^{-1}$ and increase with a factor of $\sqrt{2}$ while negative contours at shown at -3$\sigma$ (dashed line). The beam size is shown in the bottom left corner, its FWHM corresponds to 72\,arcsec.}
    \label{fig:eff}
\end{figure*}
\section{Total intensity results and analysis}
\subsection{SRT image and analysis}
Fig. \ref{fig:ciza} shows the resulting 18.6\,GHz SRT image obtaining by averaging the data between 18\,GHz and 19.2\,GHz. The noise is 0.13\,mJy\,beam$^{-1}$, the beam size is 0.9 arcmin. Solid contours are drawn from 3$\sigma$ increasing by a factor of $\sqrt{2}$, while dashed contours are the negative -3$\sigma$ contours.\\
We clearly detected the emission associated to the brightest radio sources in the field including the northern relic which extends over a length of $\sim$1.8\,Mpc with a deconvolved width ranging from $\sim$40\,kpc up to $\sim$160\,kpc. The relic surface brightness is not homogeneous across the arc, remarking the filamentary structure observed for the first time at 1.5 and 3\,GHz \citep{digennaro18}. \\
In the central part of the image, we can observe the two radio galaxies \citep[labelled D and E in Fig. 7 of][]{loi17} which are unresolved at our beam resolution. 
A patch of radio emission is located eastwards from the centre. We noticed that this structure does not correspond to the candidate relic observed at lower frequencies \citep[see][]{hoang17} and that it can be seen only in the LOFAR 145\,MHz low-resolution (i.e. 35\,arcsec) image.\\
Fig. \ref{fig:ciza_cnt_box} shows the positive contours of Fig. \ref{fig:ciza} overlaid on the WSRT+SRT 1.4\,GHz total intensity image reported in \citet{loi17}. A gray arrow indicates the eastern structure mentioned above.\\
Fig. \ref{fig:ciza_cnt_box} also shows a blue contour, traced following the 3$\sigma$ contours of the 18.6\,GHz image, which indicates the northern relic area. We measured the flux density of the relic inside this region which covers an area of 13 beam areas. Similar to what has been done in \citet{kier17}, using the same region, we evaluated the residual base level of the image by computing the mean of the surface brightness in 10 different regions of the image with no obvious source. 
The resulting mean surface brightness computed from these 10 regions is $\sim$4.5\,$\muup$Jy\,beam$^{-1}$ and the associated rms is $\sim$75\,$\muup$Jy\,beam$^{-1}$. 
The base level correction factor of the image is given by this mean multiplied by the number of beams of the relic region, and corresponds to \,58.5$\,\muup$Jy.
We compute the error on the flux density as:
\begin{equation}
   {\rm \Delta S_{\nu} = \sqrt{(f \cdot S_{\nu})^2 + \sigma^2 \cdot N_{BEAM} + (\Delta BL})^2}, 
\end{equation}
where f is the systematic flux uncertainty that we assume to be equal to 10 \%, $\sigma$ is the noise image and N$_{\rm BEAM}$ is the number of beam corresponding to the relic area. ${\rm \Delta BL}$ is the error associated to the base level correction which is equal to the rms of the regions divided by the square root of ${\rm N_{BEAM}}$. \\
The flux density of the relic at 18.6\,GHz corrected for the base level is:
\begin{equation}
    {\rm S_{18.6\,GHz}=(7.67 \pm 0.90)\,mJy.}
\end{equation}
\subsection{Effelsberg image and analysis}
Fig. \ref{fig:eff} shows the 14.25\,GHz Effelsberg image, convolved at a resolution of 72\,arcsec, shown in the bottom left corner. The noise in this image is 0.5\,mJy\,beam$^{-1}$. Solid 
contours are drawn from 3$\sigma$ increasing by a factor of $\sqrt{2}$, while dashed contours are the negative -3$\sigma$ contours.\\
In this image, we detected the central sources of the cluster, namely the D and E sources, and patches associated with the northern radio relic of CIZA J2242.8+5301. \\
Using the same blue region of Fig. \ref{fig:ciza_cnt_box}, we evaluated the relic flux in the 14.25\,GHz Effelsberg image at its original resolution of 49\,arcsec. 
At 14.25\,GHz, the relic flux is:
\begin{equation}
    {\rm S_{14.25\,GHz}=(9.5 \pm 3.9)\,mJy.}
\end{equation}

\subsection{Spectral fitting}
\begin{figure}
	\includegraphics[width=\columnwidth]{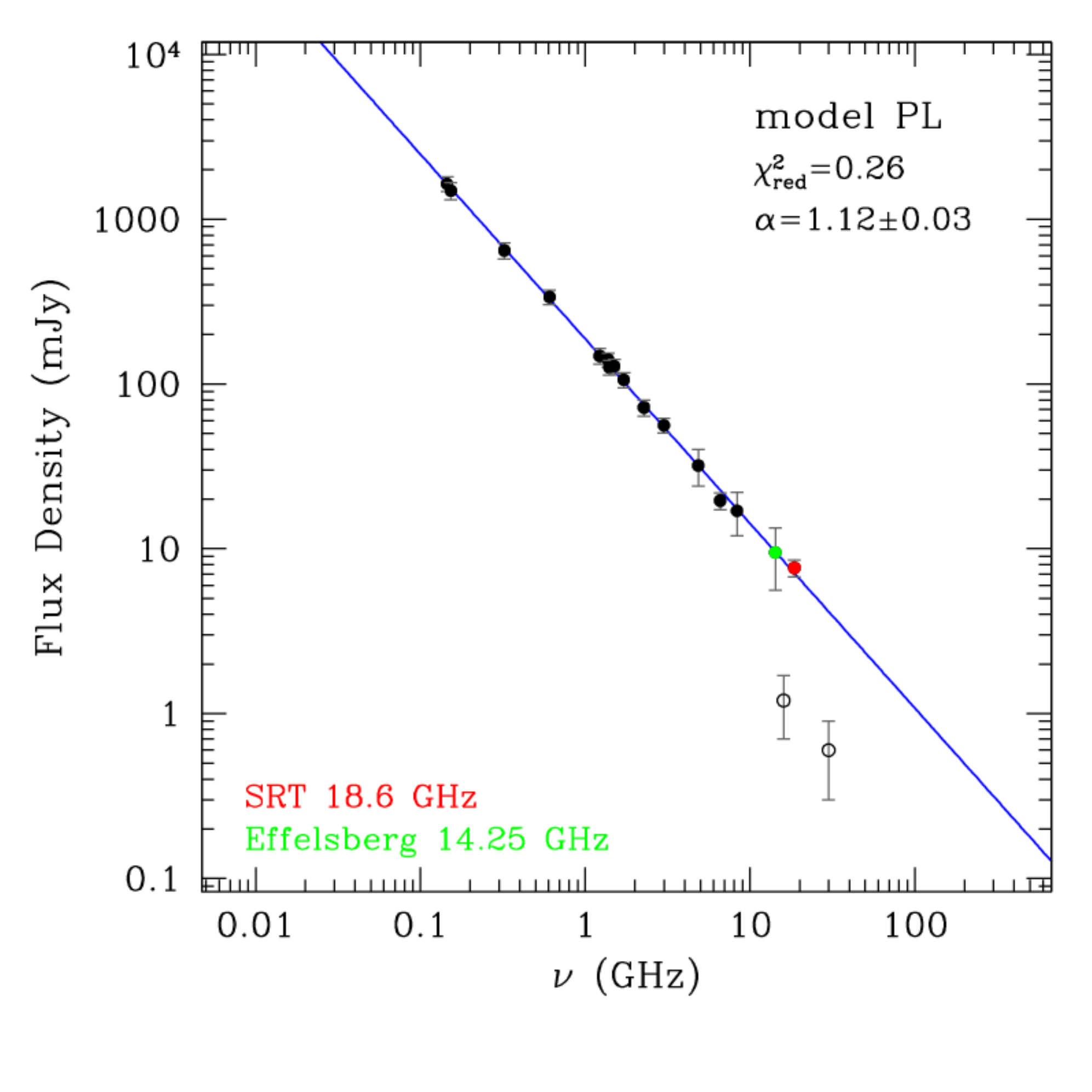}
   \caption{The CIZA J2242.8+5301 northern relic flux density as a function of frequency. Black dots are measurements from the literature, while the green and red dots are the 14.251,GHz and 18.6\,GHz measurement presented in this work. The power-law best-fit is shown as a solid blue line. The two empty points are measurements from \citet{stroe16} which we did not include in the fitting procedure.}
    \label{fig:fit}
\end{figure}

Fig. \ref{fig:fit} shows the flux density of the relic as a function of frequency. We included our 14.25\,GHz and 18.6\,GHz measurements as a green and red dot respectively to the most updated results in the literature \citep{kier17,loi17,hoang17,digennaro18} shown as black dots. More details about the measurements are in Table \ref{tab:meas}. 
\begin{table}
	\centering
	\caption{Flux density measurements of the CIZA J2242.8+5301 northern relic shown in Fig. \ref{fig:fit}. The measurement at 6.6\,GHz has been repeated considering a 3$\sigma$ threshold.}
	\label{tab:meas}
	\begin{tabular}{ccc}
		\hline
		frequency [GHz] & $S_{\nu}$ [mJy] & Reference\\
		\hline
        0.145 & 1637$\pm$168 & \citet{hoang17}\\
        0.153 & 1488$\pm$171 & \citet{hoang17}\\
        0.323 & 646$\pm$71 & \citet{hoang17}\\
        0.608 & 337$\pm$35 & \citet{hoang17}\\
        1.221 & 148$\pm$16 & \citet{hoang17}\\
        1.382 & 140$\pm$14 & \citet{hoang17}\\
        1.4   & 126$\pm$12.6 & \citet{loi17}\\
        1.5   & 128.1$\pm$12.81 & \citet{digennaro18}\\
        1.714 & 106$\pm$11 & \citet{hoang17}\\
        2.272 & 72$\pm$8  & \citet{hoang17}\\
        3     & 56.1$\pm$5.61 & \citet{digennaro18}\\
        4.85  & 32$\pm$8 & \citet{kier17}\\
        6.6   & 19.6$\pm$2.3 & \citet{loi17}*\\
        8.35  & 17$\pm$5 & \citet{kier17}\\
        14.25 & 9.5$\pm$3.9 & This work\\
        16    & 1.2$\pm$0.5 & \citet{stroe16}\\
        18.6  & 7.67$\pm$0.90 & This work\\
        30    & 0.6$\pm$0.3 & \citet{stroe16}\\
		\hline
	\end{tabular}
\end{table}
These data have been modelled with a power law spectrum which resulted to be the best-fit model, with a reduced chi-square $\chi^2_{\nu}=0.26$. In the fitting procedure, we did not include the interferometric measurements at 16 and 30\,GHz by \cite{stroe13} because we suspected that a significant fraction of the flux density could be missed in these estimates due to the lack of sensitivity at scales larger than their minimum baseline. \\
We measured again the flux at 6.6\,GHz considering now a 3$\sigma$ threshold instead of 5$\sigma$ \citep[see][for the details about this measurement]{loi17}.
The resulting integrated spectral index calculated between 145 MHz and 18.6 GHz is:
\begin{equation}
    \alpha=1.12\pm0.03,
\end{equation} 
and confirms what found in recent works over a smaller frequency range \citep[i.e. between 145\,MHz and 2.2\,GHz, and between 1.5 and 3\,GHz as reported by][respectively]{hoang17,digennaro18}. Our measurements exclude a possible steepening of the relic spectrum up to a frequency of 19\,GHz.

\section{Sunyaev-Zel'dovich decrement}
Observations at high-frequency (i.e. >10GHz) can be affected by the Sunyaev-Zel'dovich (SZ) effect which consists in a inverse-Compton interaction between cosmic microwave background (CMB) photons and thermal ICM particles. As a result, the synchrotron emission associated to cluster-embedded sources is reduced from true values because the background CMB emission is shifted towards higher frequencies \citep[see][]{birk} adding a "negative" contamination to the flux density of cluster-embedded sources.\\
We can reasonably assert that the measurements presented in this work are not affected by the large-scale radial SZ decrement since such an effect would be mitigated in the baseline subtraction procedure. Nevertheless, as described by \citet{basu}, a sharp pressure jump due to a shock is expected to generate a "localized" SZ small-scale decrement at the observed frequencies and it is reasonable that our images are affected by this contamination (even if we remark that in the case of the northern relic of CIZA J2242.8+5301, the surface brightness and pressure jump were not clearly detected \citep[but see][]{ogrean}). \\
Even if we do not have a direct measurement of the SZ at the position of the shock, we could try to estimate the expected decrement at 18.6\,GHz from numerical simulation. According to \citet{basu} the Comptonization parameter y at the relic location exceeds by -1.4$\times 10^4$ with respect to the radial y trend which we assume to have been absorbed by the baseline subtraction. Using Eq. 3 by \citet{basu} to compute the expected negative surface brightness associated to the CMB inverse-Compton emission, after multiplying this quantity by the number of beam area covered by the relic to derive the flux density $\rm S_{SZ}$, we can compute the SZ decrement at 18.6\,GHz as follows:
\begin{equation}
   \rm \Delta S_{SZ}=-2\,y\,S_{SZ}=-0.9\,mJy.
\end{equation}
This value corresponds to the uncertainty associated with our measurement. At 14.25\,GHz the decrement is equal to -0.3\,mJy. This means that it is very hard to detect a SZ decrement in our images and in fact we do not see any evident deviation from the power-law trend in the relic flux density spectrum.
We can fix the -11.7\% (which is $\rm \Delta S_{SZ}$ at 18.6\,GHz divided by the relic flux density) as an upper-limit of the contamination due to the SZ effect, assuming that no other effects are contributing to compensate it. Considering Eq. 15 by \citet{basu} reported below:
\begin{equation}
\begin{split}
\rm \bigg( \frac{S_\nu^{relic}}{\Delta S_{SZ}} \bigg) \simeq -9\cdot10^{4}
    \cdot \bigg( \frac{\xi_{e/p}}{0.05} \bigg) 
    \cdot \bigg( \frac{M}{3} \bigg) 
    \cdot \bigg( \frac{T_u}{1\,keV} \bigg)^{1/2}
    \cdot \bigg( \frac{W}{100\,kpc} \bigg)^{-1} \\
    \cdot (1+z)^{-(4+\delta/2)}
    \cdot \frac{B_{\rm relic}^{1+\delta/2}}{B_{\rm CMB}^2+B_{\rm relic}^{2}}
    \cdot\bigg( \frac{\nu}{\rm 1.4\,GHz} \bigg)^{-(2+\delta/2)},
\end{split}
\label{eq:eq}
\end{equation}
we can then tentatively investigate how the magnetic field in the relic region $\rm B_{relic}$ changes with the Mach number M assuming the previous SZ decrement upper-limit.
In the above formula, $\rm S_{\nu}^{relic}/\Delta S_{SZ}$ is the SZ decrement in percentage fixed to -11.7\%, $\xi_{e/p}$ is the ratio between relativistic protons and electron that we fix equal to 0.0026 according to \citet{kang12}, $\rm T_u$ is the upstream temperature equal to 2.7\,keV \citep{aka}, W is the relic width that is equal to $\sim$100\,kpc (see Sect. 3), $\delta$ is the spectral index of the relativistic particle distribution assumed to be equal to 4.2 \citep{kang12}, ${\rm B_{CMB}}$=3.24(1+z)$^2$\,$\muup$G.
We stress that many approximations have been done to derive this formula, for instance it is assumed that the relic is perfectly seen edge-on. However, this simplified formula can give us an alternative way to estimate of the magnetic field strength in the relic region.
\begin{figure}
	\includegraphics[width=\columnwidth]{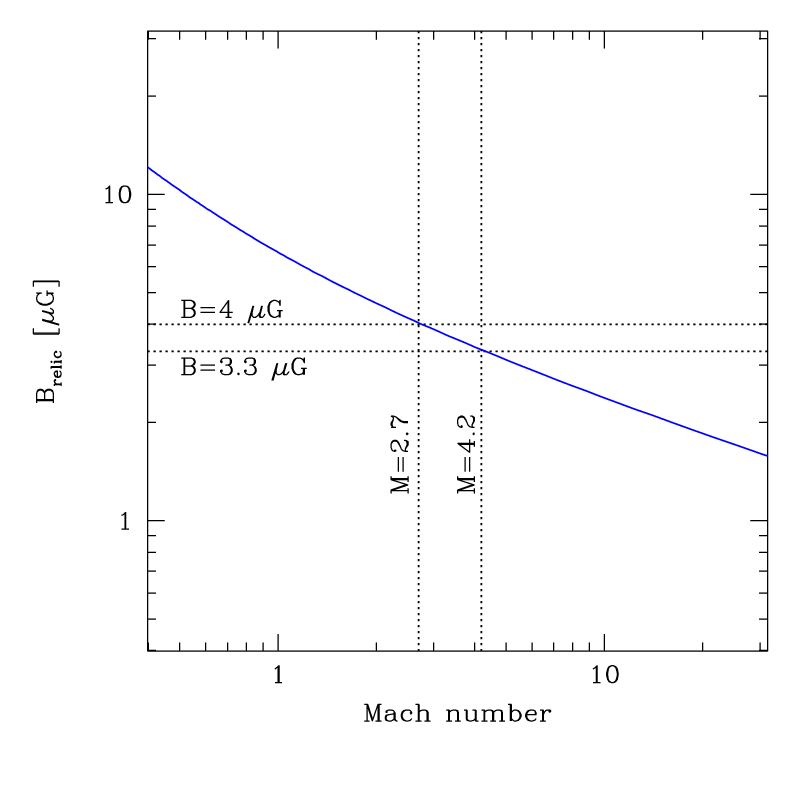}
   \caption{Magnetic field strength versus Mach number. See Eq. \ref{eq:eq}.}
    \label{fig:sz}
\end{figure}
The blue line in Fig. \ref{fig:sz} shows the relic magnetic field values at different Mach number at $\rm \nu=18.6\,GHz$. This has been derived computing the resulting Mach number values at different magnetic field. We draw dotted vertical lines at the X-ray and radio derived Mach number. To these values corresponds a magnetic field strength of $\rm B_{relic}\sim$3-4\,$\muup$G, which is an intermediate value between what found by \citet{van10}, B$\rm _{relic}\sim5-7\,\muup G$, and \citet{kier17}, B$\rm _{relic}\sim2.4\,\muup G$, by fitting the radial profile of the relic and with equipartition arguments respectively.

\section{Polarisation properties at 18.6 GHz}
\begin{figure*}
    \centering
    \includegraphics[width=1.9\columnwidth]{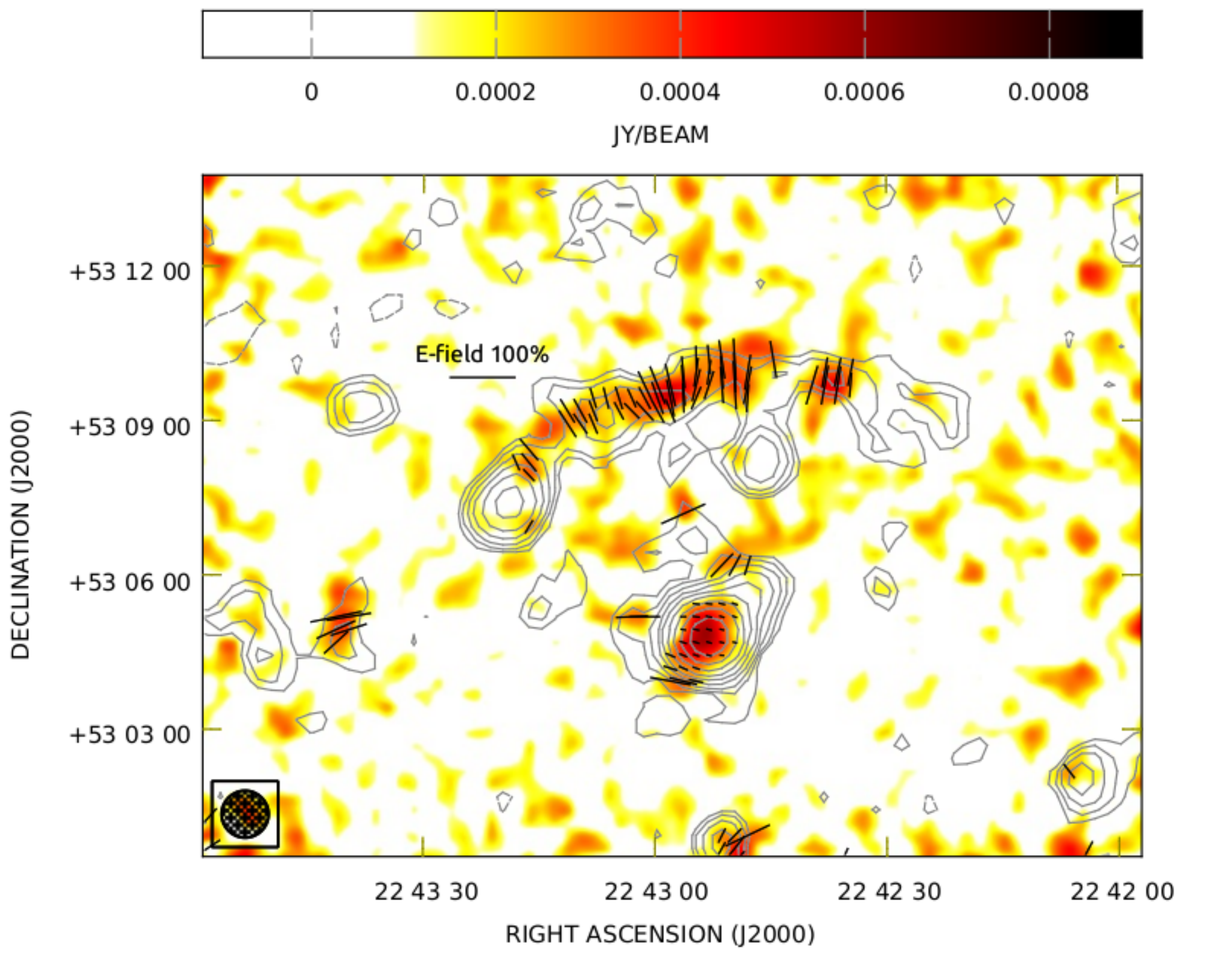}
    \caption{18.6\,GHz SRT polarised intensity image between 18\, and 19.2\,GHz. Contours are the same as the total intensity contours shown in Fig. \ref{fig:ciza}. Vectors represent the intensity and the orientation of the E-field. They have been traced for pixels with a total intensity larger than 3$\sigma$, a fractional linear polarisation SNR$\geqslant$2, and an error on the polarisation angle smaller than 15 degrees.}
    \label{fig:ciza_pol}
\end{figure*}

Fig. \ref{fig:ciza_pol} shows the polarised intensity associated with the northern and central part of the galaxy cluster CIZA J2242.8+5301 detected at 18.6\,GHz with the SRT. This is the first polarised image of a radio relic at these frequencies and it is fundamental to evaluate its intrinsic polarisation properties which are strongly related to the local magnetic field.
Contours are the total intensity contours already shown in Fig. \ref{fig:ciza}. The length and the orientation of the overlaid vectors represent the intensity and the orientation of the radio wave electric field vectors. We draw them considering all pixels with a total intensity larger than 3$\sigma$, which also have a fractional linear polarisation SNR$\geqslant$2 and errors on the polarisation angle smaller than 15 degrees.\\
\begin{figure}
    \centering
    \includegraphics[width=0.97\columnwidth]{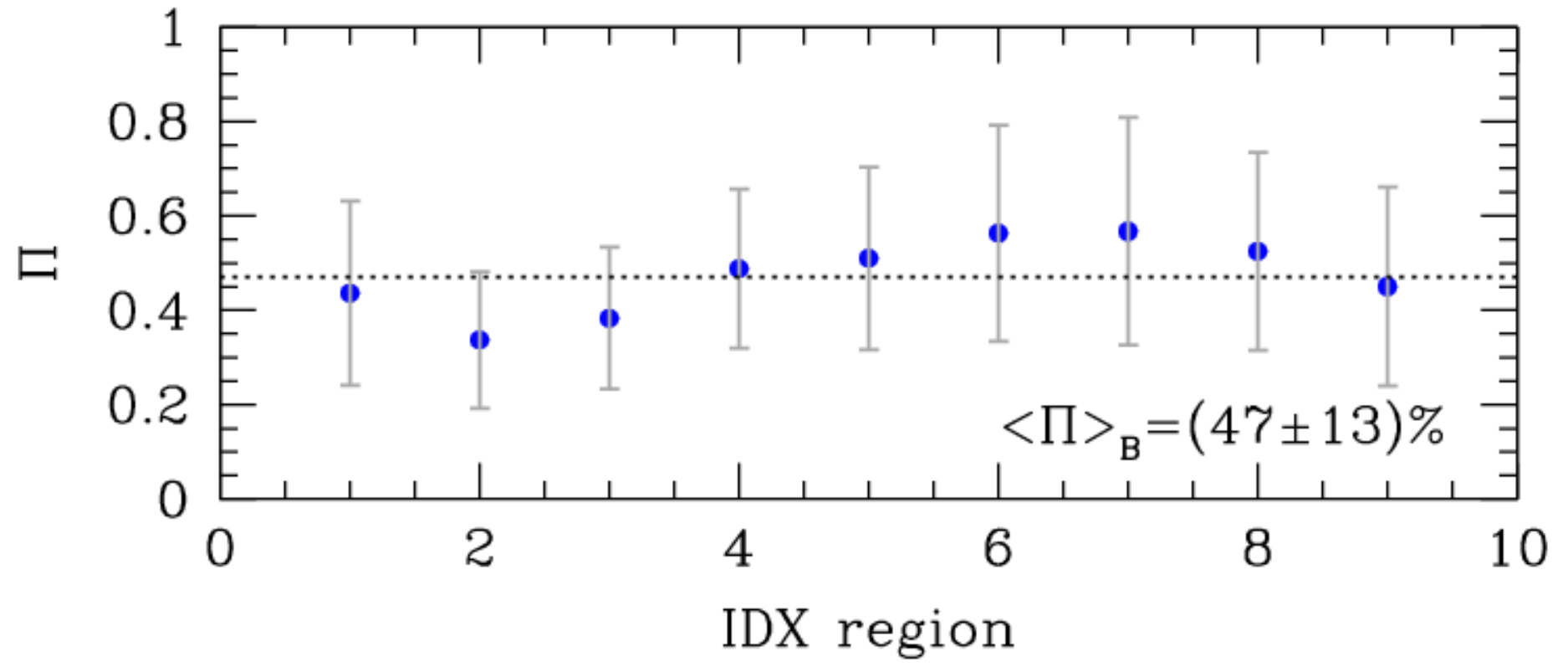}
    \caption{Polarisation fraction computed in boxes with size corresponding to the beam size. The dashed line shows the mean value reported in the bottom left corner of the plot.}
    \label{fig:ciza_fpol}
\end{figure}
The E-field vectors are aligned perpendicular to the relic filaments. This has already been observed at other frequencies, namely at 4.9\,GHz \citep{van10}, at 6.6\,GHz \citep{loi17}, at 4.85 and 8.35\,GHz \citep{kier17}, and it is expected in presence of a shock wave.
Visually, only the magnetic field lines at 90 degrees from the line-of-sight are observed, while we are blind to the other magnetic fields components, since the resulting E-vectors are not propagating towards us. The "observable" magnetic fields lines are aligned with the shock surface and the E-vectors are perpendicular to it.
We note that, if the magnetic field would already have been aligned on large scales, its non-zero components would be perpendicular to the shock direction, generating E-vectors similar to what observed in the previous case. Therefore, we cannot distinguish between a large-scale ordered magnetic field and a turbulent magnetic field structure in the presence of a shock wave which compress the magnetic field in a thin layer.\\
Fig. \ref{fig:ciza_fpol} shows the polarisation fraction computed in 9 boxes with size corresponding to the beam size located from east to west across the relic.
The average polarisation fraction at this resolution is equal to (47$\pm$13)\%.
We detected the polarised signal associated to the D and E sources at the centre of the cluster which shows a mean polarisation fraction of $\sim 12\%$. 
The eastern structure is strongly polarised, with an average polarisation fraction of $\sim$70\%. 

\section{Discussion and conclusions}
The northern relic of CIZA J2242.8+5301 is a privileged site to study the acceleration of relativistic electrons by merger shocks in the ICM. Even if it is one of the most cited radio relics, it has posed a lot of questions since its discovery, and it has challenged the standard model of shock acceleration in cluster outskirts.
With this work, we have shown that the radio relic spectral behaviour is well modelled with a power-law from 145\,MHz up to 19\,GHz. The measurements presented in this work constitute clear evidence that there is no steepening at high frequencies at variance with earlier claims \citep{stroe16}. The inconsistency between this measurement and the 16\,GHz and 30\,GHz taken with the AMI and CARMA interferometers is mostly likely due to the lack of sensitivity of these interferometers on scales larger than their minimum baseline. It is clear that interferometers can miss flux associated with sources extended on large angular scales and therefore might not be sufficient in the study of extended sources, making a combination with single-dish observations essential.\\

We know that in the context of the DSA \citep{hoeft}, from the fitted integrated spectral index, we can derive the Mach number as:
\begin{equation}
    M=\sqrt{\frac{\alpha+1}{\alpha-1}}.
\end{equation}
According to what found in this work, the Mach number should be M=4.2$^{+0.4}_{-0.6}$. This value is significantly different from what derived with X-ray observations \citep[M$_{\rm X}=2.7^{+0.7}_{-0.4}$, ][]{aka}. However, as also previous works highlighted \citep{hoang17,digennaro18}, deriving the Mach number from the integrated spectral index can lead to significant errors in some cases. Indeed, the previous formula has been derived under the assumption that the properties of the shock and the downstream gas remain constant during the electron cooling. Recent cosmological simulations of radio relics by \citet{wittor} have shown that both the Mach number and the magnetic field are not uniform across the shock front. Furthermore, they showed that the downstream magnetic field is by far not constant. By comparing spectral index measurements from observations and simulation, \citet{rajpurohit20} argued that the spectral index is most-likely biased by the high value tail of the underlying Mach number distribution.
A better way to estimate the Mach number is to evaluate it from the injected spectral index, measuring this quantity in the injection region from images at very high-resolution or from the color-color diagram in case of projection effect. \\

\begin{figure}
    \centering
    \includegraphics[width=0.47\textwidth]{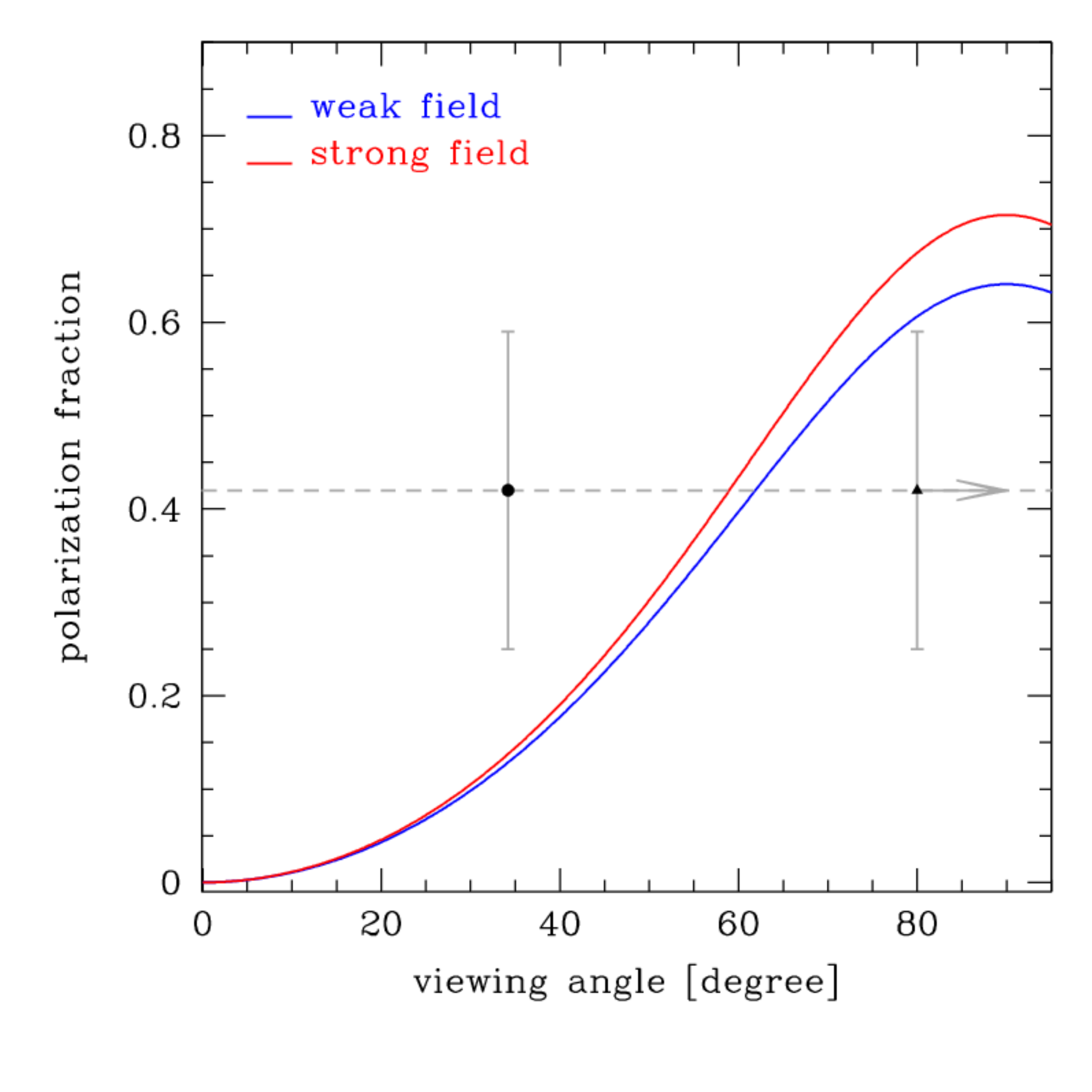}
    \caption{Polarisation fraction as a function of the viewing angle in the case of weak (blue) and strong (red) magnetic fields. }
    \label{fig:fpol_angle}
\end{figure}

Following \citet{enss98}, we can use our polarised image to investigate the possible projection of the relic in the framework of the DSA.\\
Under the assumption of a weak magnetic field in the relic region, i.e. if the magnetic pressure is lower than the gas pressure, the average fractional linear polarisation $<\Pi>$ is a function of the viewing angle $\theta$ as:
\begin{equation}
\rm <\Pi_{\rm weak}> = \frac{3\gamma+3}{3\gamma+7}\frac{\sin^2{\theta}}{\frac{2R^2}{R^2-1}-\sin^2{\theta}},
\end{equation}
where R is the compression ratio defined as:
\begin{equation}
\rm    R=\frac{\alpha+1}{\alpha-\frac{1}{2}},
\end{equation}
and $\gamma$ is the spectral index of the electron power-law like energy distribution, with $\gamma=2\alpha$+1.
On the other side, if the magnetic field pressure is higher that the gas pressure, we have:
\begin{equation}
\rm <\Pi_{\rm strong}> = \frac{3\gamma+3}{3\gamma+7}\frac{\sin^2{\theta}}{\frac{2}{15}\frac{13R-7}{R-1}-\sin^2{\theta}}.
\end{equation}
The scalar mean <$\Pi$> of the fractional polarisation is evaluated from the Q, U, and I Stokes parameters averaged over the entire relic area. If the shock relic is seen edge-on, i.e. forming a viewing angle of 90 degrees, then we observe its maximum averaged polarisation fraction, since the polarised signal is due to the amplified magnetic field vector components along the shock direction. For a relic seen face-on, the magnetic field vector components which are illuminated by the relic relativistic particles are not aligned, therefore we observe a null averaged polarisation fraction. This is shown in
Fig. \ref{fig:fpol_angle}, where we plot the polarisation fraction as a function of the viewing angle in the case of weak (blue) and strong (red) magnetic fields, assuming $\alpha=1.12$. \\
We measured the scalar mean of the polarisation fraction which is equal to <$\Pi$>=(42$\pm$17)\%. This value is slightly different from what reported in the previous section since there we measured the mean polarisation fraction from values obtained within beam size regions, while here we computed the mean from Q, U, and I averaged over the entire relic region.
Using the geometrical argument of \citet{enss98}, we computed a viewing angle of 33 degrees which means that the shock direction is forming an angle of 57 degrees with respect to the plane of the sky. A black dot shows this estimate in Fig. \ref{fig:fpol_angle}. It is clear that the assumption of a spherical geometry of the shock wave is a simplification of the real shock geometry. For this reason is not surprising that the black dot is far from the plotted curves. Nevertheless, according to the measured value of polarization fraction we can see that the radio relic is not perfectly seen edge-on, but we are observing it from a viewing angle between 45 and 80 degrees, i.e. the relic could be inclined with respect to the plane of the sky by an angle ranging from 10 to 45 degrees. 
Numerical simulation of this relic \citep{van11} determined that the shock direction is inclined with respect to the plane of the sky by an angle $\lesssim$10 degree. This estimates is shown as a black triangle in the plot.

The interpretation of the mean polarised fraction in the framework of the DSA confirms that the relic is not seen perfectly edge-on and can explain the inconsistency between X-ray and radio derived Mach number. In addition, it is important to mention that X-ray observations do not clearly confirm the presence of a shock wave, since the shock has been only observed as a jump in temperature but not in the surface brightness \citep[<2$\sigma$ detection, see][]{ogrean}. Shock waves with Mach number M$\sim$3 should show a significant jump in the X-ray surface brightness profile. \\

Another question which needs to be answered is what is the source of cosmic-ray electrons. As already pointed out in the Introduction, shock waves with weak Mach number cannot efficiently accelerate particles from the thermal pool  \citep{brunetti,van19}. Therefore, also for this particular case, we expect that a pre-existing relativistic particle population has been re-accelerated through the DSA mechanism at the shock passage. Due to the large extension of the relic, it is unlikely that the two nearby radio galaxies can supply such relativistic electrons. Nevertheless, fossil particles, injected by galaxies which are no longer active, may have accumulated in the relic area during the past history of the cluster. 
A multiple shock \citep{hong15} structure along the line-of-sight could be the responsible of the (re-)acceleration of such fossil particles in the northern relic of CIZA J2242.8+5301. This would also explain the inconsistency between radio and X-ray derived Mach numbers. Interestingly, \citet{ogrean} identified additional inner small density discontinuities both on and off the merger axis with Chandra data, which could be interpreted, as suggested by the authors, as shock fronts.\\

To summarise, we presented new measurements of the northern radio relic of CIZA J2242.8+5301 obtained with the Effelsberg and the SRT single-dish telescopes. We found a flux density of $\rm S_{14.25\,GHz}=(9.5\pm3.9)\,mJy$ and $\rm S_{18.6\,GHz}=(7.67\pm0.90)\,mJy$ at 14.25\,GHz with the Effelsberg and 18.6\,GHz with the SRT telescopes respectively. The best-fit modelling of the radio relic spectrum between 143\,MHz and 19\,GHz is a power-law with spectral index $\alpha=(1.12\pm0.03)$. Our measurements exclude a possible steepening of the relic spectrum up to a frequency of 19\,GHz. We estimated the possible SZ contamination and we determined that the expected decrement both at 14.25\,GHz and at 18.6\,GHz would be within the uncertainty associated with our measurements. Assuming the modelling of \citet{basu}, we also inferred a rough estimate of the magnetic field in the relic region based on the derived SZ decrement upper-limit at 18.6\,GHz and this resulted to be $\rm B_{relic}\sim3-4\,\muup G$, a value close to what found in previous works with different approaches \citep{van10,kier17}. For the first time, we also detected the polarised intensity associated to the relic at 18.6\,GHz. The mean polarisation fraction calculated in boxes of the same beam size of the image is equal to (47$\pm$13)\% while the scalar mean computed over the entire relic area, dividing the mean polarized intensity by the mean total intensity, is equal to (44$\pm$18)\%. In the last paragraphs, we speculated about the origin of this radio relic. We suggested that this relic emission could be due to a fossil plasma re-accelerated by a multiple shock structure which is propagating in a direction inclined with respect to the plane of the sky. High-resolution and high-frequency polarised radio images as well as deep X-ray images could help to constrain the viewing angle and the shock structure of the relic respectively, allowing us to validate the proposed scenario.

\section*{Data availability}
The data underlying this article will be shared on reasonable request to the corresponding author.

\section*{Acknowledgements}
We thank the anonymous Referee for the useful suggestions and comments which help to improve our paper.
We thank Sorina Reile for her work on reducing the Effelsberg data. FL and PS acknowledge financial support from the Italian Minister for Research and Education (MIUR), project FARE, project code R16PR59747, project name FORNAX-B. D.W. is funded by the Deutsche Forschungsgemeinschaft (DFG, German Research Foundation) - 441694982. AB acknowledges financial support from the Italian Minister for Research and Education (MIUR), project FARE. AB e MB acknowledges financial support from the ERC-Stg DRANOEL, no 714245. D.W, K.R. and F.V. acknowledge financial support from the ERC  Starting Grant "MAGCOW", no. 714196.
The Sardinia Radio Telescope \citep{bolli,prandoni} is funded by the Ministry of Education, University and Research (MIUR), Italian Space Agency (ASI), the Autonomous Region of Sardinia (RAS) and INAF itself and is operated as National Facility by the National Institute for Astrophysics (INAF). The development of the SARDARA back-end has been funded by the Autonomous Region
of Sardinia (RAS) using resources from the Regional Law 7/2007 "Promotion of the scientific research and technological innovation in Sardinia" in the context of the research project CRP-49231 (year 2011, PI Possenti): "High resolution sampling of the Universe in the radio band: an unprecedented instrument to understand the fundamental laws of the nature". Partly based on observations with the 100-m telescope of the MPIfR
(Max-Planck-Institut für Radioastronomie) at Effelsberg.






\bsp	
\label{lastpage}
\end{document}